\documentclass[prl,a4paper,showpacs,
twocolumn,
nofootinbib
]{revtex4}
\usepackage{graphicx}
\usepackage{epsfig}
\usepackage{psfig}
\usepackage{subeqn}

\def\gsim{\lower0.5ex\hbox{$\:\buildrel >\over\sim\:$}}
\def\lsim{\lower0.5ex\hbox{$\:\buildrel <\over\sim\:$}}

\newcommand{\be}{\begin{equation}}
\newcommand{\ee}{\end{equation}}
\newcommand{\beq}{\begin{eqnarray}}
\newcommand{\eeq}{\end{eqnarray}}

\def\d{\delta}

\def\l{\lambda}
\def\m{\mu}
\def\bm{\bar{\mu}}
\def\n{\nu}
\def\bn{\bar{\nu}}
\def\t{\tau}
\def\T{\theta}

\def\D{\Delta}

\begin{document}


\title{Testing whether muon neutrino flavor mixing is maximal}
\author{Sandhya Choubey}
\email{sandhya@he.sissa.it}
\affiliation{INFN, Sezione di Trieste, Trieste, Italy}
\affiliation{Scuola Internazionale Superiore di Studi Avanzati, 
I-34014 Trieste, Italy}
\vskip 0.2cm
\author{Probir Roy}
\email{probir@theory.tifr.res.in}  
\affiliation{Department of Theoretical Physics,\\
Tata Institute of Fundamental Research, 
Homi Bhaba Road, Mumbai 400 005, India}
\date{\today}

\begin{abstract} 
\vskip 0.2cm
The small difference between the survival probabilities of muon neutrino
and antineutrino beams, traveling through earth matter in a long baseline
experiment such as MINOS, is shown to be an important measure of any 
possible deviation from maximality in the flavor mixing of those states. 
\end{abstract}


\maketitle

How really close to maximal 
\cite{foot1}
is the flavor mixing of muon neutrinos and
antineutrinos discovered in atmospheric neutrino experiments
\cite{ref1}? We propose in this Letter a way of probing the deviation,
if any, from this maximality. The idea is to measure the small
difference $\D P_{\m\m} = P_{\m\m} - P_{\bm\bm}$, with $P_{\m\m} 
\equiv P \left[\n_\mu (0) \to \n_\m (L)\right]$ and $P_{\bm\bm} 
\equiv P \left[ \bn_\m (0) \to \bn_\m (L)\right]$ representing 
respective survival probabilities of muonic neutrino and antineutrino
beams after passage through a distance $L$ in earth matter. Suppose we
neglect the subdominant oscillations due to the smaller mass
difference relevant to the solar neutrino problem, but retain a
nonzero mixing between the first and the third family
neutrinos. Working to the linear perturbative order in the earth
matter effect, we are then able to show that $\D P_{\m\m}$ is
proportional to $|U_{e3}|^2 (1-2 |U_{\m 3}|^2)$, $U$ being the
Pontecorvo-Maki-Nakagawa-Sakata neutrino flavor mixing matrix, with a
computable proportionality coefficient. Thus the effect is linear in
the deviation $1/\sqrt{2} - |U_{\m 3}|$ from maximal mixing.

Much has been learnt recently about neutrino masses and mixing angles
from solar \cite{ref2}, atmospheric \cite{ref1}, reactor \cite{ref3,
ref4} and long baseline \cite{ref5} studies 
\cite{review}. We know now that the
squared mass difference between one pair of neutrinos comprising
$\n_\m$ and $\n_\t$, mixed nearly maximally $(\T_{23} \simeq
45^\circ)$
\cite{foot2}
is $|\d m^2_{32}| \sim 2 \times 10^{-3} \ eV^2$. We also
know that the squared mass difference between another pair of
neutrinos, involving $\n_e$ mixed by a large angle $(\T_{12} \simeq
30^\circ)$ with a nearly equal combination of $\n_\m$ and $\n_\t$, is
$\d m^2_{21} \sim 7 \times 10^{-5} \ eV^2$. Furthermore, the mixing
between the third possible pair, characterized by the angle
$\theta_{13}$, is known to be quite restricted, as elaborated below.

Yet longer baseline experiments with $\n, \bn$ beams/superbeams
promise to be the wave of the future in neutrino physics. MINOS
\cite{ref6} and off-axis NUMI \cite{ref7} are forthcoming experiments 
and will be followed by JPARC \cite{ref8}, CNGS \cite{ref9} and other
efforts. Many theoretical and phenomenological analyses \cite{ref10}
of physics issues pertaining to long baseline experiments have been
carried out meanwhile. However, the focus of recent studies has
largely been on appearance experiments: specifically, the `golden'
$(\n_e \to \n_\m)$ and `silver' $(\n_\m \to \n_\t)$ channels and 
their synergy in probing leptonic CP violation as well as the mixing
between the first and third family neutrinos. Less attention has been
paid to survival probabilities, the originally measured quantities
\cite{ref11} in atmospheric neutrino studies. This is since they do
not yield direct information on those aspects. However, it is the
survival probabilities for $\n_\m$ and $\bn_\m$ beams/superbeams,
specifically their difference, which should help determine the
deviation from maximality, if any, of the flavor mixing of muon
neutrinos.

Earth matter directly affects only neutrinos with electronlike
flavor. However, this gets induced 
into the other neutrino flavors indirectly
through mixing cum oscillation effects. For neutrinos of muonic
flavor, there are two sources of such an occurrence: (1) mixing between the 
electronlike flavor and the third physical neutrino through the factor
$|U_{e3}|^2$ and (2) subdominant oscillations between electronlike and
muonlike neutrino flavors, 
driven by $\d m^2_{21}$.
It is generally known, and we will show later, that the effect 
of (2) is quite small for the baseline length of MINOS \cite{ref6}.
This means that one can take the earth matter effect 
in such an experiment to be due to (1) only. We also assume that the 
actual value of $|U_{e3}|^2$ is not as small as one order of 
magnitude less than its current upper bound $0.05-0.07$ \cite{ref12} 
so that our effect will be measurable. Moreover, we shall treat this 
effect to the lowest perturbative order \cite{ref13} in $A=\sqrt{2}G_FN_e$, 
$N_e$ being the average electron density of the earth.
This will also be justified later.
On the
other hand, if $|U_{e3}|^2$ 
turns out to be 
smaller by more than
an order of magnitude
than its current upper bound, the content of our Letter will
prove empty.

The effective Hamiltonian for neutrino oscillation in matter can be
written as
\be
H = U \pmatrix{0 & 0 & 0 \cr 0 & \D_{21} & 0 \cr 0 & 0 & \D_{31}}
U^\dagger + \pmatrix{A & 0 & 0 \cr 0 & 0 & 0 \cr 0 & 0 & 0} .
\label{eq1}
\ee
Here $\D_{ij} = (m^2_i - m^2_j)/(2E_\n)$, $m_i$ and $E_\n$ being the
mass of the $i$th (physical) neutrino and the beam energy
respectively. For simplicity, let us work in the uniform earth density
approximation, though our results extend to the more general variable
density case, as shown by using the evolution operator formalism
\cite{ref14}. When $\D_{21}$ is neglected in comparison with
$\D_{31}$, the effects of the solar neutrino mixing angle $\T_{12}$
and of the CP violating phase $\d$ in $U$ become
inconsequential. Then, in the standard parametrization and with
$c_{ij} \equiv \cos\T_{ij}$, $s_{ij} \equiv \sin\T_{ij}$, $H$ takes
the form
\begin{widetext}
\beq
H &\simeq& \nonumber \\ 
&&
\!\!\!\!\!\!\!\!\!\!\!\!\!\!\!\!\!\!\!\!\!\!\! 
diag.(A,0,0) + 
\pmatrix{1 & 0 & 0 \cr 0 & c_{23} & s_{23} \cr 0 & -s_{23} & c_{23}} 
\pmatrix{c_{13} & 0 & s_{13} \cr 0 & 1 & 0 \cr -s_{13} & 0 & c_{13}} 
\pmatrix{0 & 0 & 0 \cr 0 & 0 & 0 \cr 0 & 0 & \D_{31}}
\pmatrix{c_{13} & 0 & -s_{13} \cr 0 & 1 & 0 \cr s_{13} & 0 & c_{13}}  
\pmatrix{1 & 0 & 0 \cr 0 & c_{23} & -s_{23} \cr 0 & s_{23} & c_{23}}. 
\label{eq2}
\eeq
\end{widetext}
The RHS of (\ref{eq2}) can be diagonalized
by the unitary matrix $\tilde U$, yielding the eigenvalues
$\lambda_{1,2,3}$, given by
\begin{subequations}
\beq
\l_2 &=& 0 , \\ \label{eq3b}
\l_{1,3} &=& {1\over 2} (\D_{31} + A \mp B), 
\label{eq3c}
\eeq
\end{subequations}
where 
\be
B = \sqrt{\D_{31}{}^2 + A^2 - 2\D_{31} A \cos 2 \T_{13}} .
\label{eq4}
\ee
Moreover, $\tilde U$ can be written as \cite{ref10}
\beq
\tilde U &=& \pmatrix{1 & 0 & 0 \cr 0 & c_{23} & s_{23} \cr 0 & -s_{23} &
c_{23}} \pmatrix{c_{\T_m} & 0 & s_{\T_m} \cr 0 & 1 & 0 \cr -s_{\T_m} &
0 & c_{\T_m}} 
\nonumber \\
&=& \pmatrix{c_{\T_m} & 0 & s_{\T_m} \cr -s_{23}s_{\T_m} &
c_{23} & s_{23} c_{\T_m} \cr -c_{23} s_{\T_m} & -s_{23} & c_{23} c_{\T_m}}, 
\label{eq8}
\eeq
where the angle $\theta_m$ is related to $\theta_{13}$ and $A$ by,
\beq
\tan2\theta_m = \frac{\Delta_{31}\sin2\theta_{13}}
{\Delta_{31}\cos2\theta_{13}-A}
\eeq  

The $\n_\m$ survival probability, after propagation through a distance
$L$ in matter, is \cite{ref15}
\beq
P_{\m\m} = 1&-&4 \bigg( |\tilde U_{\m 1}|^2 |\tilde U_{\m 2}|^2
\sin^2 {\l_2 - \l_1 \over
2} L 
\nonumber \\ &+& 
|\tilde U_{\m 1}|^2 |\tilde U_{\m 3}|^2 \sin^2 {\l_3-\l_1 \over
2} L \nonumber \\ 
&+& 
\ |\tilde U_{\m 2}|^2 |\tilde U_{\m 3}|^2 \sin^2 {\l_3-\l_2\over
2} L \bigg) \nonumber \\ 
&&\!\!\!\!\!\!\!\!\!\!\!\!\!\!\!\!\!\!\!\!\!= 
1 ~- ~4 \bigg(s^2_{23} c^2_{23} s^2_{\T_m} \sin^2 {\D_{31} + A - B \over 4}
L 
\nonumber \\ &+&
s^4_{23} s^2_{\T_m} c^2_{\T_m} \sin^2 {BL \over 2} \nonumber \\ 
&+& \ c^2_{23} s^2_{23} c^2_{\T_m} \sin^2
{\D_{31} + A + B \over 4} L \bigg) .
\label{eq9}
\eeq
$P_{\bm\bm}$ can be obtained from $P_{\m\m}$ simply by changing $A$ to
$-A$ \cite{ref16new}. 
Moreover, in vacuum with $A=0$ and hence $\l_1=0=\l_2$ and $\l_3
= \D_{31}$ as well as $\tilde U = U$, we have
\be
P^{\rm vac}_{\m\m} = P^{\rm vac}_{\bm\bm} = 1-4 |U_{\m 3}|^2 (1 -
|U_{\m 3}|^2) \sin^2 {\D_{31}L \over 2} .
\label{eq10}
\ee
The oscillation probability for a $\n_\m/\bn_\m$ traveling in vacuum,
namely $1-P^{\rm vac}_{\m\m}$, is maximal, corresponding to maximal
mixing, when $|U_{\m 3}| = 1/\sqrt{2}$. Though a careful measurement of
$P^{\rm vac}_{\m\m}$ may yield a value of $|U_{\m 3}|^2$ slightly
different from $1/2$, one here faces the difficulty of measuring a
small term of order $(1/\sqrt{2} - |U_{\m 3}|)^2$. 
{\it The vacuum term being quadratic in the deviation from maximality
of muon neutrino flavor mixing contrasts strikingly with the matter
effect term being linear in the said deviation.}

\begin{figure}[t]
\centerline{\epsfig{figure=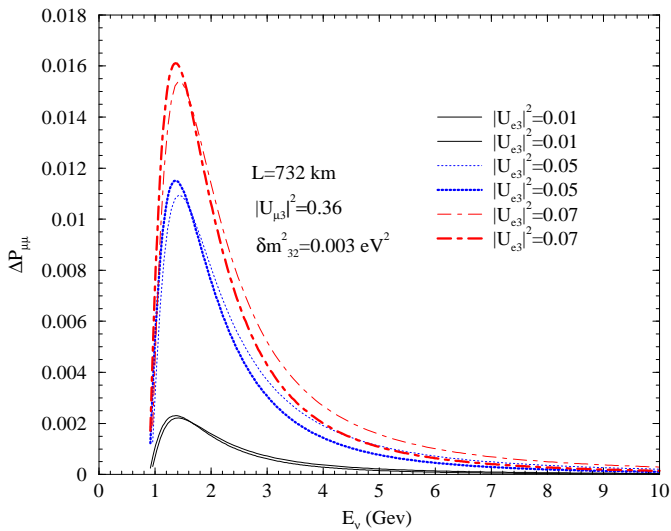,height=2.8in,width=3.5in,angle=0}}
\caption{\label{delprfig}
$\Delta P_{\mu\mu}$ as a function of 
the neutrino energy. The thick lines show the 
approximate analytic form given 
in Eq. (\ref{eq12}) 
while the thin lines are obtained by the exact numerical
solution of the equation of motion of the neutrinos traveling in 
matter. The comparison between our analytic results and the exact 
numerical solution is shown for three different values of 
$|U_{e3}|^2$.
For the exact numerical solution we have 
used $\delta m_{21}^2=7\times 10^{-5}$ 
eV$^2$ and $\sin^2\theta_{12}=0.3$.
}
\end{figure}

\begin{figure}[t]
\centerline{\epsfig{figure=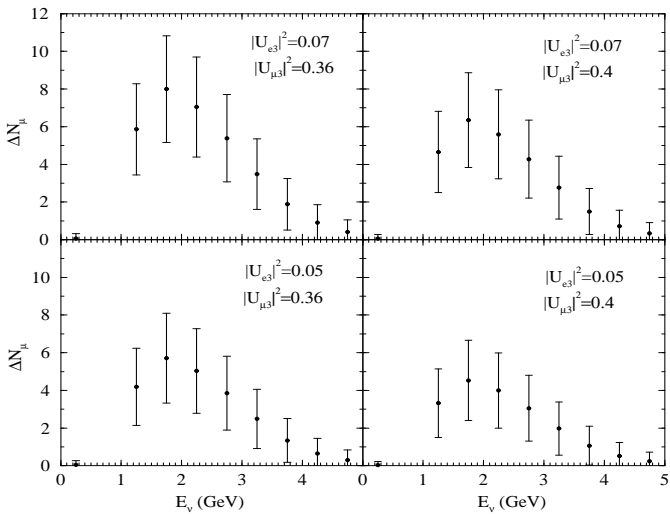,height=2.7in,width=3.5in,angle=0}}
\caption{Difference in the number of neutrino and antineutrino 
events due to the deviation of $|U_{\mu 3}^2|$ from maximality.
}
\label{delev}
\end{figure}

Let us now consider (\ref{eq9}), keeping terms only to linear order in
$A$. The $A^2$ terms 
will be shown to cancel from our effect and the corrections
will be shown to be of ${\cal O}[(U_{e3}A)^3]$. Note that, for the MINOS
baseline length and energies, $|A/ \Delta_{31}| = {\cal O}(10^{-1})$.
It is useful to note that
\begin{subequations}
\label{eq11a-e}
\beq
&&s^2_{\T_m} = s^2_{13} (1 + 2A \D^{-1}_{31} c^2_{13}) + {\cal O} (A^2) , \\
&&c^2_{\T_m} = c^2_{13} (1 - 2A \D^{-1}_{31} s^2_{13}) + {\cal O} (A^2) , \\
&&\sin^2 {\D_{31} + A - B \over 4} L = {\cal O} (A^2) , \\ 
&&\sin^2 {BL \over 2} = \sin^2 {\D_{31}L \over 2} - {AL \over 2} \cos 2 
\T_{13} \sin (\D_{31} L)  \nonumber \\
&&\qquad \qquad\qquad\qquad\qquad\qquad+ {\cal O} (A^2) , \\ 
&&\!\!\!\!\!\!\sin^2 {\D_{31} + A + B \over 4} L = \sin^2 {\D_{31} L \over 2} +
{AL \over 2} s^2_{13} \sin (\D_{31} L) \nonumber \\
&&\qquad \qquad\qquad\qquad\qquad\qquad+ {\cal O} (A^2).
\eeq
\end{subequations}
Utilizing (9a-e) in (\ref{eq9}), we obtain 
\beq
P_{\m\m} &=& P^{\rm vac}_{\m\m} + 2 s^2_{23} c^2_{13}s^2_{13} 
(c^2_{23} - s^2_{23} \cos 2 \T_{13} ) \nonumber \\ [2mm]
&&  A \left[ 4\D^{-1}_{31} \sin^2 {\D_{31}L \over 2} 
- L\sin (\D_{31}L)\right]  + {\cal O} (A^2) \nonumber \\[2mm] 
&=& P^{\rm vac}_{\m\m} + 2 |U_{e3}|^2 |U_{\m 3}|^2 (1-2|U_{\m
3}|^2) \nonumber \\ [2mm]
&& A \left[4 \D^{-1}_{31} \sin^2 {\D_{31}L
\over 2}-L\sin (\D_{31}L) \right]  + {\cal O} (A^2), \nonumber
\eeq
leading to
\beq
\D P_{\m\m} &=& 
4 |U_{e3}|^2 |U_{\m 3}|^2 
(1-2 |U_{\m 3}|^2) \nonumber \\ [2mm]
&& \!\!\!\!\!\!\!\!\!\!\!\!
A \left[4 \D^{-1}_{31} \sin^2
{\D_{31}L \over 2}- L \sin (\D_{31}L) \right]  
\nonumber \\ && \!\!\!\!\!\!\!\!\!\!\!\!+ {\cal O}[( U_{e3} A)^3] .
\label{eq12}
\eeq

Eq.(\ref{eq12}) is the key result. It shows that the linear $A$ term
in $\D P_{\m\m}$ is proportional to $1- 2|U_{\m 3}|^2$ which is the
deviation from maximality of the $\n_\m$ flavor 
mixing in vacuum, as evident
from (\ref{eq10}). We may also note that 
the ${\cal O} (A^2)$ terms cancel between $P_{\m\m}$ and $P_{\bm\bm}$.
Moreover, the corrections involve ${\cal O}[( U_{e3} A)^3]$ terms 
since the lowest order $A-$terms in the transition amplitude 
\cite{ref17} come with coefficients $|U_{e3}|^2$ and $U_{e3}$.
Thus the $A^3$ terms are further suppressed.

Our two significant approximations, i.e. ignoring $\delta m_{21}^2$-driven
subdominant oscillations and the $O[(U_{e3}A)^3]$ terms in 
(\ref{eq12}), were for the convenience of analytical calculations.
A numerical code has been developed \cite{ref18} to calculate 
$\Delta P_{\mu\mu}$, treating both the subdominant and the 
earth matter effect exactly. In Fig. \ref{delprfig} the thin 
(thick) lines show the $\Delta P_{\mu\mu}$ of exact numerical 
evaluation (of (\ref{eq12})) against $E_\nu$ with the other 
parameters specified in the figure labels. The differences are quite small,
thus enhancing our confidence in the use of (\ref{eq12}) as a measure 
of $|U_{e3}|^2(1-2 |U_{\m 3}|^2) $. Interestingly, even for 
$|U_{e3}|^2=0.01$ when the magnitude of $\Delta P_{\mu\mu}$ is a lot 
less, these differences are small. 
The explanation why subdominant oscillations are
suppressed is that the $A \Delta_{21}$ term in the transition amplitude
[20] comes with a coefficient proportional to $U_{e3}$.
We have also checked numerically that, while the location
(in $E_{\nu}$) of the $\Delta P_{\mu\mu}$ peak is sensitive to variations
in $\delta m^2_{31}$, {\it its magnitude is not}.

The sensitivity of MINOS to $\Delta P_{\mu\mu}$ is best discussed in 
terms of $\Delta N_\mu$, the difference  -- due to the deviation 
of $|U_{\mu 3}|^2$ from maximality -- between the number of expected 
neutrino and antineutrino events. For an anticipated MINOS 
$\nu_\mu$ exposure \cite{ref19} of $16\times 10^{20}$ primary 
protons on target (p.o.t), this has been plotted in Fig. \ref{delev} 
along with the $1\sigma$ statistical errorbars for the parameter values 
shown in the labels. Given the detection cross-section for $\bar\nu_\mu$'s 
to be about half that of $\nu_\mu$'s, we have assumed twice as much 
exposure for $\bar\nu_\mu$'s as for $\nu_\mu$'s. The plot demonstrates
the feasibility of such a measurement.


In conclusion, we have shown how the measurement of $\D P_{\m\m}$, in
a long baseline experiment such as MINOS, can probe the deviation 
$1/\sqrt{2} - |U_{\m 3}|$ from maximality of the flavor mixing 
of the muon neutrino. The exclusion (with errors taken into account) 
of a vanishing value of $\Delta P_{\mu\mu}$ will simultaneously 
demonstrate $U_{e3}\neq 0 \neq \frac{1}{\sqrt{2}}-|U_{\mu 3}|$.
The accuracy 
in the determination of
$|U_{e3}|(\frac{1}{\sqrt{2}}-|U_{\mu 3}|)$
will depend 
more sensitively on the measurement error in
$\Delta P_{\mu\mu}$ than
on the then uncertainty in the knowledge 
of $\delta m^2_{31}$.

We thank Yuval Grossman, Sanjib Mishra, Hitoshi Murayma, Sandip
Pakvasa and Serguey Petcov for helpful discussions and Michele Frigerio 
for correcting an error.



\begin{thebibliography}{99}
\bibitem{foot1}
In a three-generation framework, 
which is essential to our analysis,
one can define maximal
flavor mixing for the muon neutrino as the situation when, 
traveling in vacuum, it converts
its flavor with the maximum probability. If the latter is calculated
neglecting the subdominant
oscillation due to $\delta m^2_{21}$, as done in Ref. \cite{ref1}, 
one finds $|U_{\mu 3}| = 1/\sqrt{2}$, see
our later discussion after Eq. (\ref{eq10}).
Now if $|U_{e 3}|$ is $\epsilon, |U_{\tau 3}| = 1/\sqrt{2}  (1 - 2
\epsilon^2)^{1/2}$.
\bibitem{ref1}
Super-Kamiokande collaboration, Y. Hayato, talk given at the EPS 2003
conference (Aachen, Germany, 2003),
{\it http://eps2003.physik.rwth-aachen.de}. S. Fukuda et al.,
Phys. Lett. {\bf B539}, 179 (2002).
\bibitem{ref2}
SNO collaboration, S.N. Ahmed et al., nucl-ex/0309004.
\bibitem{ref3}
CHOOZ collaboration, M. Appolonio et al., Eur. Phys. J {\bf C27}, 331
(2003); see also the website 
{\it http://www.nu.to.infn.it/exp/all/chooz} for their current 
statement.    
\bibitem{ref4}
KamLAND collaboration, K. Eguchi et al., Phys. Rev. Lett. {\bf 90},
021802 (2003).
\bibitem{ref5}
K2K collaboration, I. Kato, talk given at the 38th {\it Recontres de
Mariond on Electroweak Interactions and Unified Theories} (Les Ares,
France, 2003), hep-ex/0306043.
\bibitem{review}
M.~C.~Gonzalez-Garcia and Y.~Nir,
Rev.\ Mod.\ Phys.\  {\bf 75}, 345 (2003);
%
S.~Pakvasa and J.~W.~Valle,
arXiv:hep-ph/0301061;
%
V.~Barger, D.~Marfatia and K.~Whisnant,
arXiv:hep-ph/0308123.
%
%
\bibitem{foot2}
More precisely, $\sin^2 2\theta_{23} >  0.92$ at 90\% C.L. 
\cite{ref1}.
\bibitem{ref6}
MINOS collaboration, R. Saakian, Nucl. Phys. Proc. Suppl. {\bf 111},
169 (2002). M.V. Diwan, ibid. {\it al.} 172 {\bf},
272 (2003).
\bibitem{ref7}
D. Ayres et al., hep-ex/0210005. 
\bibitem{ref8}
Y. Itow et al., hep-ex/0106019. 
\bibitem{ref9}
F. Arneodo, talk given at the TAUP 2003 conference,
{\it http://mocha.phys.washington.edu/taup2003/};
K. Kodama, talk given at the Nufact 2003 conference,
{\it http://www.cap.bnl.gov/nufact03/agenda\_ug1.xhtml}.
\bibitem{ref10}
The following is a partial list. A. de Rujula, M.B. Gavela and P. 
Hernandez,
Nucl. Phys. {\bf B547}, 21 (1999). K. Dick, M. Freund, M. Lindner and
A. Romanino, Nucl. Phys. {\bf B562}, 29 (1999). V. Barger, S. Geer and
K. Whisnant, Phys. Rev. {\bf D61}, 053004 (2000). A. Bueno,
M. Campanelli and A. Rubbia, Nucl. Phys. {\bf B573}, 27
(2000);
 M.~Freund, M.~Lindner, S.~T.~Petcov and A.~Romanino,
Nucl.\ Phys.\ B {\bf 578}, 27 (2000);
A. Cervera et al., Nucl. Phys. {\bf B608}, 301 (2001); 
{\it errtm. ibid} {\bf B593}, 731 (2001). P. Huber, M. Lindner and W. 
Winter, Nucl. Phys. {\bf B645}, 3 (2002). M.V. Diwan et al., Phys. 
Rev. {\bf D68}, 012002 (2003). D. Autiero et al., hep-ph/0305185. 
O.L.G. Peres and A. Yu. Smirnov, hep-ph/0309312.
\bibitem{ref11}
T. Kajita and Y. Totsuka, Rev. Mod. Phys. {\bf 73}, 85 (2001). 
\bibitem{ref12}
One needs to be prudent at this point in quoting an absolute upper
bound on $|U_{e3}|^2$. The latter depends crucially on the lower bound
on $|\d m^2_{32}|$ coming from the Super-K atmospheric neutrino studies
\cite{ref1}. The 90\% C.L. bound $s^2_{13} < 0.04$, claimed by the
CHOOZ collaboration \cite{ref3}, is valid only for $|\d m^2_{32}| > 2
\times 10^{-3} {\rm eV}^2$; it increases sharply for lower $|\d
m^2_{32}|$ on account of reduced sensitivity. In order to derive a
reliable upper bound on $|U_{e3}|^2$, covering the entire Super-K 90\% C.L.
mass squared range $1.3 \times 10^{-3} {\rm eV}^2 < |\d m^2_{32}| <
3.1 \times 10^{-3} {\rm eV}^2$, one needs to resort to a global
analysis of neutrino oscillation data. However, such a procedure needs
to input the precise $|\d m^2_{32}|$ range currently reported by
Super-K.
A cautious upper bound on $|U_{e3}|^2$ derived from such
analyses is $\sim 0.05-0.07$, but it increases to 0.096 if only CHOOZ data
are used, see for eg.,
A. Bandyopadhyay, S. Choubey, S. Goswami, S.T. Petcov and
D.P. Roy, hep-ph/0309174; G.L. Fogli, E. Lisi, A. Marrone,
D. Montanind, A. Palazzo and A.M. Rotunno, 
Phys. Rev D {\bf 69}, 017301 (2004); 
M. Maltoni, T, Scwetz, M. A. Tortolla and J. W. F. Valle, Phys. Rev. 
D  {\bf 68}, 113010 (2003).
\bibitem{ref13}
J. Arafune, M. Koike and J. Sato, Phys. Rev. {\bf D56}, 3093 (1997);
{\it errtm. ibid} {\bf D60}, 119905 (1999).
\bibitem{ref14}
B. Brahmachari, S. Choubey and P. Roy, Nucl. Phys. B671 {\bf},483 (2003),
cf. eq.(30).
\bibitem{ref15}
S.M. Bilenky and S.T. Petcov, Rev. Mod. Phys. {\bf 59}, 671 (1989). 
\bibitem{ref16new}
For a $\nu_\mu~(\bar\nu_\mu)$ beam, $A$ is proportional to the threshold 
four-fermion limit for the 
charged current part of the forward $\nu_e e \rightarrow   \nu_e e~
(\bar\nu_e e \rightarrow   \bar\nu_e e)$ scattering amplitude and these 
differ only by a sign. 
\bibitem{ref17}
The transition amplitude $S_{\mu\mu}(L)$, to the lowest order in 
$A$ and $\Delta_{21}$, can be calculated to be $1-2i|U_{\mu 3}|^2
e^{-i\Delta_{31}L/2}\sin(\Delta_{31}L/2)-i\Delta_{21}L|U_{\mu 2}|^2
-2A|U_{\mu 3}|^2|U_{e3}|^2[e^{-i\Delta_{31}L/2}\sin(\Delta_{31}L/2)
(L-2i\Delta_{31}^{-1})-iL]+
A\Delta_{21}Re(U_{\mu 2}U_{e2}^* U_{e3}U_{\mu 3}^*)
[L^2 + 2iL\Delta_{31}^{-1}-4i\Delta_{31}^{-2}
e^{-i\Delta_{31}L/2}\sin(\Delta_{31}L/2)]$.
\bibitem{ref18}
This has been done in collaboration with S. Goswami.
\bibitem{ref19}
{\it http://www-numi.fnal.gov/}.
\end{thebibliography}
\end{document}